\documentclass[aps,pra,twocolumn,showpacs,groupedaddress,preprintnumbers]{revtex4}
\usepackage{epsfig}
\usepackage{longtable}
\usepackage{amsfonts}

\newcommand{\n}{{\vec n}}

\begin{document}



\title{Electromagnetic cavities and Lorentz invariance violation}
\author{H. M\"{u}ller, C. Braxmaier, S. Herrmann, and A. Peters}
\affiliation{Institut f\"{u}r Physik, Humboldt-Universit\"{a}t zu Berlin, 10117 Berlin,
Germany.\\ Tel. +49 (30) 2093-4907, Fax +49 (30) 2093-4718}
\email{holger.mueller@physik.hu-berlin.de}
\homepage{http://www.uni-konstanz.de/quantum-optics}
\author{C. L\"{a}mmerzahl}
\affiliation{Institute for Experimental Physics, Heinrich--Heine--University D\"{u}sseldorf, D-40225 D\"{u}sseldorf}
\affiliation{accepted for publication in Phys. Rev. D (2003)}


\begin{abstract}
Within the model of a Lorentz violating extension of the Maxwell sector of the standard
model, modified light propagation leads to a change of the resonance frequency of an
electromagnetic cavity, allowing cavity tests of Lorentz violation. However, the
frequency is also affected by a material-dependent length change of the cavity due to a
modified Coulomb potential arising from the same Lorentz violation as well. We derive the
frequency change of the cavity taking both into account. The new effects derived are
negligible for present experiments, but will be more pronounced in future tests using
novel resonator materials.
\end{abstract}


\maketitle


\section{Introduction}\label{introduction}

Einstein's special relativity and the underlying principle of Lorentz invariance are
among the foundations of modern physics. In the past, they have been tested frequently,
with increasing precision. Among others, optical means (e.g., \cite{MM,KT,BrilletHall})
are used for such tests. Today, space-borne instrumentation (SUMO \cite{SUMO} and OPTIS
\cite{OPTIS}, for example) as well as terrestrial experiments
\cite{Braxmaier,MuellerASTROD,Wolf} using ultrahigh precision optical and microwave
techniques are performed or proposed to explore the validity of this fundamental theory
even further. One reason to continue testing SR is because it is one of the pillars of
modern physics. Another reason is that most approaches towards a quantum theory of
gravity such as string theory and loop gravity predict Lorentz violation at some level
\cite{KosteleckySSB,ellis,gambini,alfaro}.

Electromagnetic radiation has provided the first glimpse of Lorentz invariance in the
famous Michelson-Morley experiment \cite{MM,MuellerASTROD}, establishing the direction
invariance of the speed of light $c$. Kennedy-Thorndike tests \cite{KT,Braxmaier}
establish the invariance of $c$ under observer Lorentz boosts. These experiments have
been performed with increasing accuracy, today using electromagnetic cavities instead of
interferometers. They are based on the modification of $c$ arising from Lorentz violation
(depending on the observer frame, direction of propagation, and polarization), which in
turn changes the resonance frequency $\nu = n c/ (2L)$ of a cavity of the Fabry-Perot
type, where $n=1,2,3, \ldots$ and $L$ is the cavity length. (This applies for cavities
where the radiation propagates in vacuum, i.e., the index of refraction is 1.) A change
of $\nu$ can be detected sensitively due to the potentially very high frequency stability
$\delta \nu/\nu\sim 10^{-15}$ possible with cavities.

Tests of Lorentz invariance can be analyzed within an extension of the standard model
developed by Kosteleck\'{y} and co-workers \cite{KosteleckySME,Kostelecky2002}. The classical
test theories, such as the Robertson-Mansouri-Sexl test theory
\cite{Robertson,MS,Kintest} or the $c^2$ formalism \cite{Will93}, can be recovered as
special cases of this model \cite{Kostelecky2002}. The standard model extension starts
from a Lagrangian formulation of the standard model, adding all possible Lorentz
violating terms that can be formed from the known particles together with Lorentz
violation parameters that form Lorentz tensors. In the purely electromagnetic sector,
these parameters are given by a tensor $(k_{F})_{\kappa \lambda\mu\nu}$ which has 19
independent components \cite{Kostelecky2002,comment2}. These Lorentz violation parameters
can be viewed as remnants of Planck-scale physics of an underlying fundamental theory,
such as string theory, at an attainable energy scale. Cavity tests are interesting, since
they can in principle access all 19 components of $(k_{F})_{\kappa
\lambda \mu \nu}$. The relative change of $\nu$ is calculated in \cite{Kostelecky2002} assuming
$L=$const, so this can be interpreted as a modification of the phase velocity of light in
vacuum $c=c_0+\delta c$ with \cite{comment1}
\begin{equation}\label{deltac}
\frac{\delta c}{c_0}=\frac 12 [(\hat N \times
\hat E^*)(\kappa_{HB})_{\rm lab}(\hat N \times  \hat E) -\hat E^* (\kappa_{DE})_{\rm lab} \hat
E] \, .
\end{equation}
Here, $\hat N$ is a unit vector pointing along the length of the cavity and $\hat E$ is a
unit vector vector perpendicular to $\hat N$ that specifies the polarization (Fig.
\ref{Ringmode}). The asterisk denotes complex conjugation. $(\kappa_{HB})_{\rm lab}$ and
$(\kappa_{DE})_{\rm lab}$ are $3\times 3$ matrices resulting from a decomposition of
$(k_{F})_{\kappa \lambda \mu \nu}$ in the laboratory frame \cite{Kostelecky2002}.
Comparison of Eq. (\ref{deltac}) to the experiments leads to upper limits on some of the
$(k_{F})_{\kappa \lambda \mu \nu}$ in the range of $10^{-8...-15}$. Because these values
are small, here and throughout the paper it is sufficient to work to lowest order in the
components of $(k_{F})_{\kappa \lambda \mu \nu}$.

However, a modification of the Coulomb potential $\Phi(\vec x)$ arising from Lorentz
violation in general affects the length $L$ of the cavity and thus the cavity resonance
frequency. This has already been pointed out in the literature
\cite{Will92,Will93,LaemmerzahlHaugan01,Kostelecky2002}. Since $\delta \nu/\nu=\delta c/c-\delta
L/L$, a length change $\delta L$ connected to $\Phi(\vec x)$ might alter or even
compensate the Lorentz-violating signal of the experiments. Owing to the different
electromagnetic composition of different solids, $\delta L$ will depend on the material.
Indeed, very early Michelson-Morley experiments have been repeated using exotic materials
(pine wood and sandstone) for the interferometer to exclude that an accidental
cancellation was responsible for the once puzzling null result \cite{MorleyMiller}.

Within the standard model extension, modifications of $c$ as well as $\Phi(\vec x)=e^2
/(4\pi |\vec x|)+V(\vec x)$ can be treated on a common basis. For many-particle systems,
like solids,
\begin{equation}\label{coulomb}
V=\frac 12 \frac{e^2}{8\pi} {\sum_{ab}}' e_ae_b \frac{\vec x_{ab} \cdot
\kappa_{DE} \cdot \vec x_{ab}}{|\vec x_{ab}|^3} \, .
\end{equation}
\cite{Kostelecky2002}. Here, $e$ is the
electron's charge and $e_a$ an integer, so that $ee_a$ is the charge of the $a$th
particle. The vector $\vec x_{ab}$ is the displacement between the $a$th and the $b$th
charge. The factor $1/2$ corrects the double-counting of pairs, the prime denotes that
$a=b$ is excluded from the summation. This equation shows that in the case of a modified
$c$ as given by Eq. (\ref{deltac}), modifications of the Coulomb potential $\Phi(\vec x)$
{\em necessarily} arise. This is due to the covariance of the Maxwell equation, i.e., a
modified wave equation implies a modified Poisson equation.

In this work, the influence of this modified Coulomb potential to the length of the
cavity is derived. As a result, we find a somewhat increased sensitivity of experiments
to some of the parameters $(k_{F})_{\kappa
\lambda \mu \nu}$. That increase is not more than a few percent for materials commonly used (quartz and
sapphire), but more pronounced for ionic materials. As a main conclusion, we thus confirm
that it is possible to neglect effects due to Lorentz violation in the Coulomb potential
in the interpretation of present cavity tests of SR.

In Sec. \ref{general}, we derive the length change $\delta L$ for ionic crystals, giving
explicit numbers for cubic crystals in Sec. \ref{cubic}. The case of covalent bonds is
discussed in Sec. \ref{discussion}. The result is discussed in \ref{conclusion}. In the
Appendix, we show that in our model, the repulsive potential between the atoms in the
lattice is almost not changed due to Lorentz violation, so we need to consider only the
modification of the Coulomb potential.

\section{Length change of crystals}

\subsection{General considerations}\label{general}

The hypothetical Lorentz violating length change of a cavity depends on the material it
is made of. As a first model, we will treat ionic crystals; we discuss the generalization
of the result to covalent bonds later.

A lattice is formed so that equilibrium between the attractive and repulsive forces
between the atoms is reached. In ionic crystals, attraction is caused by Coulomb
interactions between the ions. The repulsive potential is due to the overlap of their
orbitals. Both the attractive and the repulsive potential may be changed due to Lorentz
violation. In the appendix, we will find that the modification of the repulsive potential
is negligible compared to the modification of the attractive potential, so that it is
possible to assume non modified repulsive forces for the rest of this paper.

For this analysis we consider a cube, which is small compared to the dimensions of the
solid under consideration, but large compared to the lattice, so that boundary terms can
be neglected. Without Lorentz violation, the cube's side-length is $L$; Lorentz violation
changes this to $L+\delta L^i$ (the superscript $i=1,2,3$ denotes the spatial components
in the laboratory frame), thus distorting the cube. The solid consists of a number of
these cubes, and the fractional change $\delta L^i/L$ of the solid's length are the same
as the fractional change for a single cube. It is thus sufficient to consider a cube of
side-length $L$. The total energy of the lattice depending on $\delta L^i$ can be written
as
\begin{equation}\label{etotal}
E(L+\delta L^i) = \mbox{const}+\frac 12 E_{\rm Y}L(\delta L^i)^2+
\frac{\partial V}{\partial L^i}\delta L^i
\end{equation}
where $E_{\rm Y}$ is the Young modulus \cite{Woan}. The term proportional to $E_{\rm Y}$
is Hooke's law (for simplicity, we restrict ourselves to isotropic elasticity). Whereas
$L$ minimizes $E(L)$ without Lorentz violation, $L+\delta L^i$ is obtained by minimizing
the total energy including $V$ as given by Eq. (\ref{coulomb}). In that equation, the
summation $\sum_{ab}$ can be replaced by summing over $a$ only and multiplying with the
total number of interacting charges $N$, i.e. the number of ions in an ionic lattice.
Denoting $(x^i)_{ab}$ the $i$-th spatial component of $\vec x_{ab}$, and $l^i_j$ the
vector components of the primitive translations $\vec l_j$, we have $(x^i)_{ab}=n^jl_j^i$
with a vector $\vec n=(n^1, n^2, n^3) \in
\mathbb{R}^3$; thus
\begin{equation}
V=\frac 12 \frac{e^2}{8\pi}N (\kappa_{DE})_{ij} l^i_k l^j_l {\sum_{\vec n}}' e_{\vec n}
\frac{n^k n^l }{(n^q l^n_q n^p l^n_p)^{3/2}} \, .
\end{equation}
The summation is carried out over all ions. For computing the length change, we need the
derivative
\begin{eqnarray}\label{derivative}
\frac{\partial{V}}{\partial{l^n_m}} & = &
\frac 12 \frac{Ne^2}{8\pi}[(\kappa_{DE})_{nj}l^j_l s^{ml}+(\kappa_{DE})_{in}l^i_k s^{km} \nonumber \\
& & - 3(\kappa_{DE})_{ij}l_k^i l_l^j l_o^m t^{klno}] \, ,
\end{eqnarray}
where
\begin{equation}
s^{km}={\sum_\n}' e_\n \frac{n^kn^m}{(n^pl_p^i n^q l_q^i)^{3/2}}
\end{equation}
and
\begin{equation}
t^{klno}={\sum_\n}' e_\n \frac{n^kn^ln^nn^o}{(n^pl_p^i n^q l_q^i)^{5/2}} \, .
\end{equation}
We have $s^{nl}=s^{ln}$ and $t^{klno}$ is invariant under permutations of all indices.

The length $L$ of the sample is a linear combination of the primitive translations
$l^i_j$. Thus, the derivatives $\partial l^n_m / \partial L_i$ can be obtained. We
express
\begin{equation}
\frac{\partial V}{\partial L_i}=\frac{\partial V}{\partial l_m^n }
\frac{\partial l^n_m}{\partial L_i} \, .
\end{equation}
and insert this into Eq. (\ref{etotal}). From that, $\delta L_i$ can be obtained by
minimizing the energy $E$ given by Eq. (\ref{etotal}). We will do that explicitly for a
cubic crystal in the next section.

\subsection{Cubic symmetry}\label{cubic}

To give an explicit result we consider a crystal that without Lorentz violation has cubic
symmetry (NaCl structure). Lorentz violation will in general break this symmetry. For the
crystal without Lorentz violation, $(l_i^j)=a \delta_i^j$ with a lattice constant $a$.
Since next neighbors have opposite charges, $e_\n=(-1)^{n^1+n^2+n^3}$, we find
\begin{equation}
s^{ab}=\frac{1}{a^3} \sum_\n  \frac{(-1)^{n_1+n_2+n_3}n_1^2}{(n^k n^k)^{3/2}}
\delta^{ab} =: \sigma \frac{1}{a^3}\delta^{ab} \, .
\end{equation}
The summation is carried out over $\{\mathbb{Z}^3 \backslash 0\}$. $t^{abcd}$ is only
nonzero if the indices are two pairs of equal indices. Thus,
$t^{aabb}=t^{abab}=t^{abba}=: t^{ab}$ with
\begin{equation}
t^{ab}=\left\{ \begin{array}{lr} \frac{1}{a^5}\sum_\n
\frac{(-1)^{n_1+n_2+n_3}n_1^2n_2^2}{(n^k n^k)^{5/2}}=:\tau_{\perp} \frac{1}{a^5}\, , & a\neq
b\, ,
\\  \\ \frac{1}{a^5} \sum_\n \frac{(-1)^{n_1+n_2+n_3}n_1^4}{(n^k  n^k)^{5/2}}=:\tau_{\|} \frac{1}{a^5}\, , & a=b \, . \end{array}
\right.
\end{equation}
Numerical summation (summing from $-50 \ldots 50$) gives $\sigma \approx -0.58,
\tau_{\perp} \approx 0.23$, and $\tau_{\|}=-1.04$. Using these equations, we obtain (no
summation over $n$)
\begin{equation}
\frac{\partial V}{\partial
l_n^n }=\frac 12
\frac{Ne^2}{8\pi}[2a\sigma (\kappa_{DE})_{nn}-3a^3(\kappa_{DE})_{aa}t^{an}]
\end{equation}
Let the $z$ axis be parallel to $\hat N$ [as it was defined below Eq. (\ref{deltac})].
Thus, we have
\begin{equation}
\frac{\partial V}{\partial L_z}=\frac{\partial V}{\partial l_z^z}\frac{\partial l_z^z}{\partial
L_z}=\frac{\partial V}{\partial l_z^z}\frac{1}{\eta_z}
\end{equation}
where the (constant) $\eta_z=L/a$ is the number of unit cells making up $L$ with the
lattice parameter $a$. Setting to zero the derivative of Eq. (\ref{etotal}) we obtain the
fractional length change in $z$ direction. Substituting
\begin{equation}
N= N_m \frac{N_A V \rho}{M} \, ,
\end{equation}
where $N_A$ is Avogadro's constant, $V$ the crystal's volume, $\rho$ its density, $M$ its
molecular weight, and the factor $N_m$ is the number of atoms (ions, in an ionic crystal)
per molecule, we obtain
\begin{eqnarray} \label{lengthchange}
\frac{\delta L_z}{L} & = & A
\left[(2\sigma-3\tau_\|) (\kappa_{DE})_{\|}-3\tau_{\perp}(\kappa_{DE})_\perp \right]
\end{eqnarray}
with $(\kappa_{DE})_\|=(\kappa_{DE})_{zz}$,
$(\kappa_{DE})_\perp=(\kappa_{DE})_{xx}+(\kappa_{DE})_{yy}$ and
\begin{equation}\label{afaktor}
A=-\frac 12 \frac{e^2\iota^2 v_1 v_2}{8\pi E_{\rm Y}} \frac{N_m N_A \rho}{M a}
\, .
\end{equation}
Here, we included a dimensionless factor $\iota^2$, which measures the effective charge
of the ions. For most crystals, $\iota=1$, for non ionic materials, $0< \iota < 1$ (see
Sec. \ref{discussion}). We also included factors $v_1$ and $v_2$, the number of valence
charges for the atoms. Using the dreibein $\hat N, \hat E, \hat N \times \hat E$, this
gives
\begin{eqnarray}\label{deltal}
\frac{\delta L}{L} & = & a_\|\hat N(\kappa_{DE})_{\rm lab} \hat N \\
&  & + a_\perp[\hat E^* (\kappa_{DE})_{\rm lab} \hat E +(\hat N \times \hat E^*)
(\kappa_{DE})_{\rm lab} (\hat N
\times \hat E)] \, . \nonumber
\end{eqnarray}
Here, $a_\|=A(2\sigma-3\tau_\|)$ and $a_\perp=-3A\tau_{\perp}$. This is now a vector
equation which will hold in any coordinates (this does not mean that the cavity length
depends on the polarization). For this equation, it is not necessary to make some
assumption about the orientation of the cavity axis relative to the lattice, since
cubical lattices are isotropic. For lattices of lower symmetry, the parameters in Eq.
(\ref{deltal}) would, however, become dependent on the angle between the crystal axes and
the cavity orientation $\hat N$.

In practical experiments, the material might not be a single crystal, but a non
crystalline material. These consist of a large number of microscopic crystals that are
randomly oriented. If these microscopic crystals are cubical, however, Eq. (\ref{deltal})
will hold for any of them as well as for the macroscopic body, since in this case the
equation makes no reference to the orientation of the lattice.

The coefficients $a_\|, a_\perp$, and $A$ for some materials are given in Table
\ref{Avalues}. Without Lorentz violation, the ionic model predicts the correct
distance of ions within a few percent error. Thus, we can expect a similar precision for
the Lorentz-violating length change of ionic materials. At present, such materials are
uncommon for cavities in high precision physics, but their use has been proposed
\cite{BraxmaierPRD} because of the availability of ultra-pure specimens.

Materials of other than cubic structure will show a length change of similar magnitude:
The structure enters this computation via the constants $\sigma$ and $\tau$. These are in
analogy to the Madelung-constants $\alpha$ of solid-state physics, which are of order
unity and do not depend very much on the actual crystal structure (see e.g.
\cite{Ashcroft} for some values). This also holds for $\sigma$ and $\tau$. However, lattices
with lower symmetry will lead to a more complicated expression than Eq.
(\ref{lengthchange}), in which also non diagonal terms of $(\kappa_{DE})_{ij}$ enter.\\

\subsubsection*{Length change for microwave resonators}
Sometimes, especially in microwave experiments, cavities are used where the
electromagnetic radiation of the cavity mode goes along the circumference of an annulus
or a sphere. For cubic crystals, it is straightforward to generalize our result to a
cavity of any geometry. Therefore, we parametrize the closed path of the mode by means of
a parameter $s$. We denote the length of the resonator mode path $L_m$. Its relative
change is given by integrating Eq. (\ref{deltal}) along the closed path of the mode:
\begin{eqnarray}\label{deltalring}
\frac{\delta L_m}{L_m} & = &  \frac 1S \oint  [ a_\|\hat N(\kappa_{DE})_{\rm lab} \hat N  +
a_\perp \hat E^* (\kappa_{DE})_{\rm lab} \hat E \nonumber \\ & & + a_\perp (\hat N \times
\hat E^*) (\kappa_{DE})_{\rm lab} (\hat N
\times \hat E)]\,ds
\end{eqnarray}
with $S=\oint ds$. Here, the dreibein $\hat N, \hat E$, and $\hat N \times \hat E$ is
used, where $\hat N$ gives the local orientation of the wave vector at each value of $s$
and $\hat E \perp \hat N$ gives the polarization. This equation holds for cubic lattices
(that have no preferred axis); it also holds for uniaxial crystals (such as sapphire, for
example), as long as the mode is orthogonal to the crystal axis for all values of the
parameter $s$. In this case, however, the values of $a_\perp$ and $a_\|$ have to be
computed specifically for the uniaxial crystal.

\begin{figure}[t]
\centering
\epsfig{file=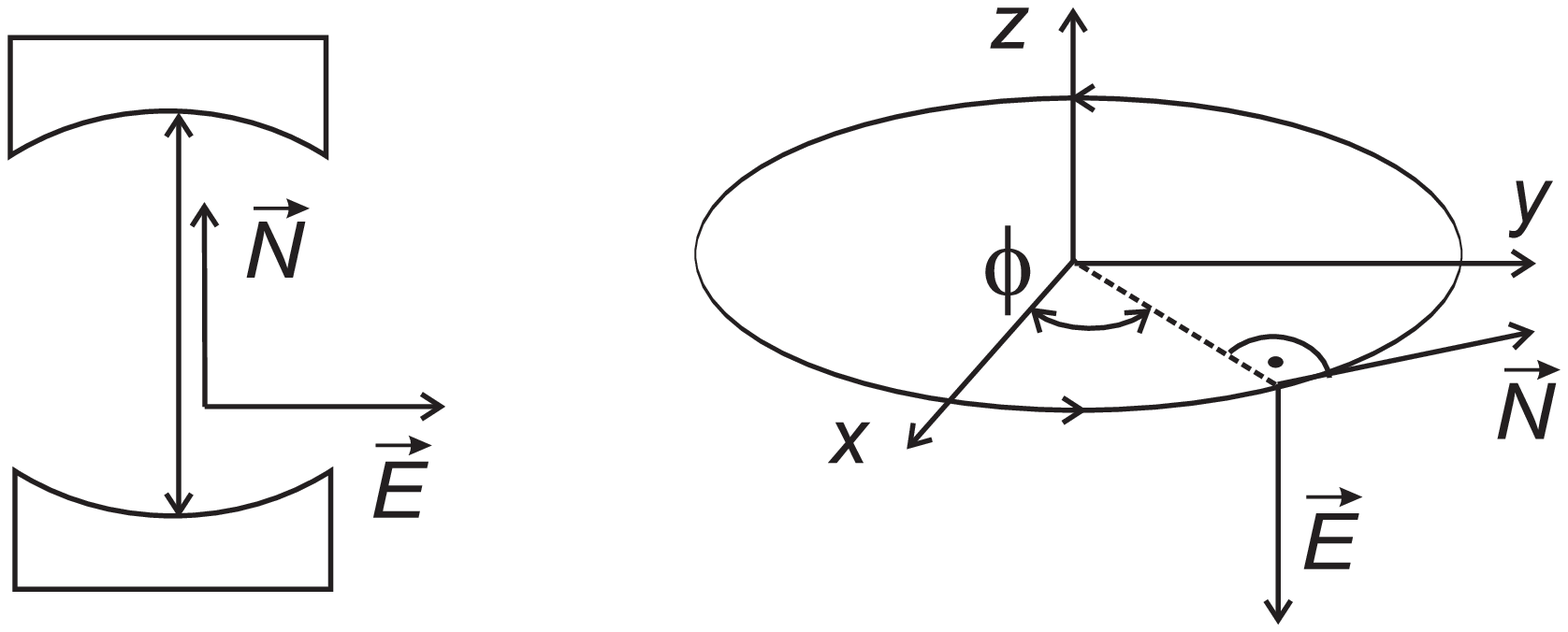, width=0.45\textwidth}
\caption{Left: Optical (Fabry-Perot) cavity. The unit vector $\hat N$ is parallel to the
cavity axis, the unit vector $\hat E$ is the polarization of the electromagnetic
radiation inside the cavity. Right: Microwave cavity. The ring denotes the
electromagnetic cavity mode, parametrized by an angle $0 \leq \phi< 2\pi $. The
polarization $\hat E$ is parallel to the $z$ axis, the unit vector $\hat N$ is a tangent
denoting the direction of the mode for each $\phi$.
\label{Ringmode}}
\end{figure}

For example, for a cavity where the mode has the geometry of a ring with the $z$ axis
being the symmetry axis, we identify the parameter $s$ as the angle $0 \leq
\phi< 2\pi$ (Fig.
\ref{Ringmode}); we have $\hat N = (\cos
\phi, \sin \phi, 0)$. Choosing $\hat E= (0,0,1)$ parallel to the $z$ direction,
\begin{equation}
\frac{\delta L_m}{L_m}= \frac{a_\|+a_\perp}{2}[(\kappa_{DE})_{xx}+(\kappa_{DE})_{yy}]+a_\perp
(\kappa_{DE})_{zz}
\end{equation}
for the path length of a microwave cavity mode like it is shown in Fig. \ref{Ringmode}.
This result is valid for a cavity made from a cubic crystal; for a uniaxial crystal, it
is valid as long as the crystal $c$ axis is parallel to the $z$ axis (provided that
$a_\perp$ and $a_\|$ are computed for the particular material).

In whispering gallery resonators, a dependency of the index of refraction on Lorentz
violation would also have to be taken into account. Compared to the length change, this
probably is a small correction, since in whispering gallery resonators only a part of the
electromagnetic field energy travels within the material.

\subsection{Covalent bonds}\label{discussion}

The materials quartz (used, for example, in \cite{BrilletHall,HilsHall}) and sapphire
(e.g., \cite{Braxmaier,COREs,MuellerASTROD}) are based on covalent bonds. These bonds
have a partial ionic character, i.e., the effective charges of the atoms are $\iota e v$
(as before, $v$ denotes the number of valence charges), where $\iota<1$ can be determined
roughly from the difference of the electronegativities of the atoms (see, e.g.,
\cite{Mortimer}). From Eq. (\ref{afaktor}), we thus obtain
length-change coefficients in the per cent range (Table \ref{Avalues}). Since the concept
of partial ionic character is not an exact concept, these values have limited accuracy.
They suggest, however, that the relative length change $\delta L/L$ of a cavity made from
quartz or sapphire is negligible compared to the relative change $\delta c/c$ of the
speed of light.

Here, we give some arguments indicating that the length change of a covalent crystal may
be derived in analogy to ionic crystals. Covalent bonds are given by Coulomb interactions
between delocalized electrons, and between the electrons and the atom cores. For our
calculation, the main difference to ionic bonds is that the electrons are not localized;
instead, they are described by a symmetrized $n$-electron state $\psi^n(\vec x_1, \vec
x_2, \ldots \vec x_n)$. This state is formed from the single-electron states $\psi_i$ as
the Slater determinant $(1/\sqrt{N!})\,$Det$(\psi_i(\vec x_j))$. The covalent potential
$V_C=V_{EC}+V_{EE}+V_{CC}$ is then the sum of the core-core interactions $V_{CC}$
\begin{equation}
V_{CC} = \sum_a \sum _b \Phi^{(ab)}(\vec x_a-\vec x_b)
\end{equation}
[where $a$ and $b$ enumerate the cores, and $\Phi^{(ab)}(\vec x_a-\vec x_b)$ is the
Coulomb potential between the cores $a$ and $b$ at the locations $\vec x_a$ and $\vec
x_b$], the electron-core interactions
\begin{equation}
V_{EC}=\sum_i
\sum_a
\int (\psi^n)^{*}
\Phi^{(ia)} (\vec x_i - \vec x_a) \psi^n d^{3} x_1 \ldots d^3 x_n
\end{equation}
(where $i$ numerates the electrons and $\Phi^{(ia)}$ is the Coulomb potential between the
core $a$ and the electron $i$) and the electron-electron interactions
\begin{equation}
V_{EE}=\sum_i \sum_j \int (\psi^n)^{*}
\Phi^{(ij)} (\vec x_i - \vec x_j) \psi^n d^{3} x_1 \ldots d^3 x_n
\end{equation}
with $\Phi^{(ij)}$, the Coulomb potential between the electrons $i$ and $j$. Since
$\psi^n$ is symmetrized, these expressions contain all the terms usually found in the
treatment of covalent systems, like the hydrogen molecule. Although the electron
wave-functions are spread over the whole crystal volume, their centers of mass are
localized at some point within a distance $\bar a$ from the cores, where $\bar a$ is of
atomic dimensions
\cite{kohn}, smaller than the typical distance of atoms in a lattice. As a toy model, we
thus assume centers of charge at points $\vec x_{\bar a}$ at some distance $\bar a$ from
the locations of the nearest cores. Coulomb forces between these centers of charge and
the cores make up an important part of the covalent binding force. The model reduces
$V_{EC}$ and $V_{EE}$ to a sum over Coulomb interactions, i.e., we have
\begin{eqnarray}
V_{EC} & \sim & \sum_a \sum _{\bar a} \Phi^{(\bar a a)} (\vec x_{\bar a} - \vec x_{a})
\\ V_{EE} & \sim &
 \sum _{\bar a}
\sum _{\bar b} \Phi^{(\bar a \bar b)} (\vec x_{\bar a} - \vec x_{\bar b}) \, .
\end{eqnarray}
The distance $\bar a$ between the centers of mass of the delocalized electrons and the
cores gives rise to an effective charge of the cores. A calculation of the
Lorentz-violating length change could now start from these sums and proceed as for ionic
crystals. While this is a simplified model, it indicates that the effect of a Lorentz
violating Coulomb potential for covalent bonds will be in analogy to ionic crystals.
Thus, estimates for covalent crystals made above should at least give the correct order
of magnitude.

\section{Conclusion and Summary}\label{conclusion}

\begin{table}[t]
\caption{Length change coefficients. The
first part of the table contains cubical two-atom crystals, for which the calculation was
performed. The length-change coefficients $A$, $a_\|$, and $a_\perp$ are obtained from
Eqs. (\ref{lengthchange}) and (\ref{afaktor}). The second part contains crystals of other
structure with non ionic bindings. Quartz and sapphire have two lattice constants; the
axis of highest symmetry is given. \label{Avalues}}

\begin{tabular}{|lcccccccc|}

\tableline

Material & $E_{\rm Y}$ & $\rho$ & $M$ & $\iota^2v_1v_2$ & $a$ & $A$ & $a_\|$ & $a_\perp$
\\  & (GPa) & $\left( \frac{\mbox{g}}{\mbox{cm}^3} \right) $ & & & (\AA ) & & & \\ \hline NaCl & 40 & 2.1 & 58
& 0.64 & 2.841 & -0.14 & -0.28 & 0.10
\\ LiF & 64.8 & 2.64 & 26 & 1 & 2.015 & -0.54 & -1.06 & 0.37
\\ NaF & 64.8 & 2.73 & 32 & 1 & 2.135 & -0.43 & -0.83 & 0.29
\\  \hline sapphire & 497 & 4.0 & 102 & $1.5$ & $13.0$ & -0.02 & -0.03 & 0.01 \\
quartz & 107 & 2.2 & 60 & $0.72$ & 5.4 & -0.06 & -0.11 & 0.04 \\ \hline
\end{tabular}
\end{table}

The contributions from Eq. (\ref{deltac}) and Eq. (\ref{deltal}) give
\begin{eqnarray}\label{deltanu}
\frac{\delta \nu}{\nu} & = & -a_\| \hat N (\kappa_{DE})_{\rm lab} \hat N \nonumber\\ & &
- \left(\frac12+a_\perp \right) \left[\hat E^*(\kappa_{DE})_{\rm lab}\hat E\right.  \\ &
& +
\left. (\hat N\times \hat E^*)(\kappa_{DE})_{\rm lab}(\hat N\times
\hat E)  \right] \nonumber
\, .
\end{eqnarray}
This has been simplified by noting that astrophysical tests lead to
$(\kappa_{HB})=-(\kappa_{DE})$ with an accuracy many orders of magnitude higher than
attainable in cavity experiments \cite{Kostelecky2002}. The second part of Eq.
(\ref{deltanu}) is equivalent to the speed of light change Eq. (\ref{deltac}) times $1+2
a_\perp$, $a_\perp \sim 0.01 \ldots 0.4$ for the materials in table
\ref{Avalues}. Additionally, the frequency is sensitive to a term proportional to $a_\|
\sim -0.03 \ldots -1$. For both coefficients, the small coefficients in the per-cent range
are for covalent materials (quartz and sapphire); although they are based on a simplified
model, they should be accurate enough to conclude that the relative length change of a
cavity made from quartz or sapphire is negligible against the change of the speed of
light. Since present experiments use these materials, it is indeed possible to analyze
them as if the cavity length would not be affected by Lorentz violation. This holds for
cavities of any structure, including (microwave) cavities using radial or whispering
gallery modes. Their length change is given by Eq. (\ref{deltalring}). However, future
experiments using resonators out of ionic materials, as proposed in
\cite{BraxmaierPRD} will be affected more strongly by the Lorentz violating Coulomb
forces (see Table \ref{Avalues}); here, it is necessary to analyze them using Eq.
(\ref{deltanu}) rather than Eq. (\ref{deltac}).

We note that a cavity experiment is mainly sensitive to the parity odd coefficients
\cite{Kostelecky2002} of the standard model extension, since both the change of the phase
velocity of light, Eq. (\ref{deltac}), as well as the cavity length change Eq.
(\ref{deltal}) are dominated by these coefficients. Parity even coefficients that are not
tightly constrained by astrophysical birefringence measurements \cite{Kostelecky2002}
enter cavity experiments only suppressed by $\beta_{\rm lab}$, the velocity of the
laboratory with respect to the frame in which one defines the parameters $\kappa_{DE}$
and $\kappa_{HB}$. For experiments on Earth and using a sun centered frame, $\beta_{\rm
lab} \sim 10^{-4}$.\\

We have calculated the length change of electromagnetic cavities due to a modified
Coulomb potential arising necessarily from a Lorentz violating velocity of light, as it
follows from an extension of the standard model. As a first model, we have assumed that
the cavity is made of an ionic crystal. We then extended the model to the covalent bonds
of practical materials using the approximate concept of partial charges. Taking this into
account for the interpretation of cavity tests of Lorentz invariance, we have shown that
the length change effect adds to the hypothetical Lorentz-violating signal derived from a
change of the speed of light and increases somewhat the sensitivity of experiments.
However, for materials used in present experiments, the increase is at the percent level
and thus negligible, as all the Lorentz invariance tests are null tests. As a result, the
classical interpretation that neglects the length change is justified (this also holds
for microwave cavities and for the classic interferometer tests). For future experiments
using ionic crystals as a resonator material, however, the effects derived here are
larger and must be taken into account.\\

As an outlook, we note that the properties of matter and thus the dimensions of cavities
also depend on Lorentz-violating terms in the fermionic equations of motion, i.e., a
modified Dirac equation \cite{KosteleckySME,Laemmerzahl98}. Strong upper limits on some
of these terms have been placed by, e.g., comparisons between atomic clocks
\cite{Kostelecky99}. In this work, we have focused on the purely electromagnetic sector;
the effect of the fermionic terms in cavity experiments will be treated elsewhere. It
might be possible to access Lorentz violating fermionic terms in cavity experiments due
to the length change they cause. One would compare cavities made from different
materials, in order to separate the electromagnetic from the fermionic terms.

\acknowledgements
We are grateful to V.A. Kosteleck\'{y} for discussions. It is a pleasure to acknowledge the
ongoing cooperation with S. Schiller.

\appendix

\section*{Appendix: The repulsive force of the ionic bond}

Here, we calculate the modification of the repulsive force due to Lorentz violation. It
is caused by the modified wave functions of the ions arising from the modified Coulomb
potential between the nucleus and the outer electrons. The repulsive force is a
short-range force, so it is sufficient to consider next neighbors only.

Since in ionic crystals, the positive ions are usually small compared to the negative
ones \cite{Mortimer}, we assume as a model that the positive ions are pointlike. The
repulsive potential
\begin{equation}\label{BM}
\Phi_{BM}(\vec R)=\mbox{const} |\psi^{(0)}(\vec R)|^2
\end{equation}
is then proportional to the probability density of the outer electrons of the negative
ions. $\psi^{(0)}(\vec R)$ is the unperturbed wave function at the location $\vec R$ of
the positive ion without Lorentz violation (we assume that the ion-ion interaction does
not change appreciably the wave function, which should be satisfied to reasonable
accuracy in ionic crystals). The wave function $\psi^{(0)}(\vec R)$ is proportional to
$e^{-\zeta |\vec R|}$ for large $|\vec R|$ with a constant $\zeta \sim 1/a_0$ with the
Bohr radius $a_0$; the potential $\Phi_{BM}$ is thus a Born-Mayer potential $Be^{-2 \zeta
|\vec r|}$ ($B$ as well as $\zeta$ are phenomenological parameters that are determined by
fitting the model to the measurements), which is frequently used as a model of the
repulsive force
\cite{Ashcroft}. In the case of Lorentz violation, $\psi(\vec r)=\psi^{(0)}+\psi^{(1)}$
with a correction $\psi^{(1)}$ that is proportional to $\kappa_{DE}$, which will be
calculated for hydrogenlike atoms.

\begin{figure}[t]
\centering
\epsfig{file=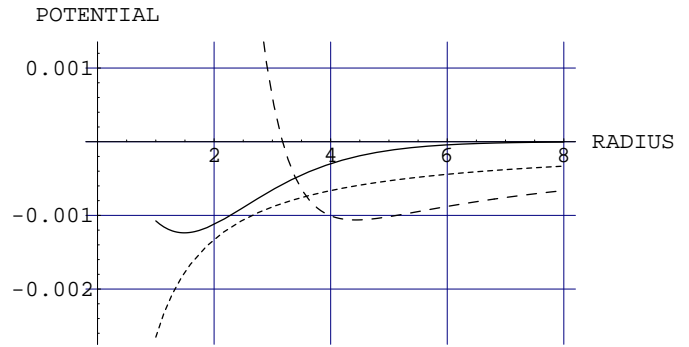, width=0.5\textwidth}
\caption{Ion-ion potentials for $n=1, l=0, m=0$ (for other quantum numbers, a similar
situation is found). The $x$ axis gives $R$ in units of $a_0$, the $y$ axis the potential
in arbitrary units. Solid line: $300 \delta \Phi_{BM}(\vec R)$, i.e. the correction to
the Born-Mayer Potential, assuming $(\kappa_{DE})^{zz}=1$, enlarged by a factor of 300.
Dots: $V(\vec R)$, i.e. the correction to the Coulomb interaction between two ions for
comparison. Dashes: Unperturbed ion-ion Potential $\Phi_{BM}-1/(4\pi |R|)$. The constant
const$=15$ has been inserted as to obtain a minimum of the total unperturbed ion-ion
potential around $R$=5, as commonly found in ionic crystals (e.g., $R=5.3a_0$ for NaCl).
The correction to the repulsive potential is negligible compared to the correction of the
Coulomb potential.
\label{potentialvergleich}}
\end{figure}

For this calculation, we denote the unperturbed state of hydrogen $|n l m>^{(0)}$ with a
principal quantum number $n$, angular momentum $l$ and magnetic quantum number $m$. The
corresponding wave function $\psi_{nlm}^{(0)}(\vec r)$ can be factorized into a radial
function $R_{nl}(r)$ that depends solely on the radius coordinate $r$, and spherical
harmonics $Y_l^m(\theta, \phi)$. Only the ground state $n=1$ is not degenerate; for the
$n=2$ states, the use of the perturbation theory of nondegenerate states is still
possible since, due to the nature of the matrix elements calculated below, the degenerate
$n=2$ states do not mix. For $n=1,2$, the state $|n l m>=|n l m>^{(0)}+|n l m>^{(1)}$ in
first order perturbation theory is given by
\begin{equation}
|n l m>^{(1)}={\sum_{n'l'm'}}' \frac{<nlm|V|n'l'm'>}{E_n-E_{n'}}|n'l'm'>
\end{equation}
with $V$ given by Eq. (\ref{coulomb}). Inserted into Eq. (\ref{BM}), this leads to a
correction to the Born-Mayer potential
\begin{eqnarray}
\delta \Phi_{BM}(\vec R)=2\, \mbox{const}{\sum_{n'l'm'}}' \frac{<nlm|V|n'l'm'>}{E_n-E_{n'}} \\
\times |\psi_{nlm}^{(0)}(\vec R)\psi_{n'l'm'}^{(0)}(\vec R)| \,. \nonumber
\end{eqnarray}
We apply coordinates in which $\kappa_{DE}$ is diagonal with the diagonal elements
$\kappa_{xx},
\kappa_{yy},$ and $\kappa_{zz}$. Applying polar coordinates with the $\theta=0$ axis
parallel to the $z$ axis, the Lorentz violating correction to the Coulomb potential $V$
can be written as
\begin{equation}
V=\frac{e^2}{8\pi r}[(\kappa_{xx} \sin^2 \phi+\kappa_{yy} \cos^2 \phi)\sin^2
\theta+\kappa_{zz}\cos^2 \theta]\, .
\end{equation}
Since in first order perturbation theory, the change of the wave function $|nlm>^{(1)}$
for a sum of perturbations $\sum_i V_i$ is a linear combination of the changes for each
single perturbation $V_i$, it is sufficient to consider the term proportional to
$\kappa_{zz}$. In configuration space,
\begin{eqnarray}
< nlm |r^{-1}\cos^2 \theta | n'l'm'>= \int R_{nl}(r)\frac 1r R_{n'l'}(r)r^2dr \nonumber
\\
\times \int Y_l^m(\theta, \phi) \cos^2 \theta Y_{l'}^{m'} (\theta,\phi)\sin \theta d\theta d\phi
\quad
\end{eqnarray}
with the normalized radial functions
\begin{equation}
R_{nl}=\frac{2}{n^2}\sqrt{\frac{\zeta^3(n-l-1)!}{(n+1)!}}\left(2\frac{\zeta r}{n}
\right)^l e^{-\zeta r/n}L_{n-l-1}^{2l+1}\left(2\frac{\zeta r}{n} \right)
\end{equation}
where $L_p^k(z)$ are associated Laguerre polynomials as defined in \cite{Gradstein}. For
the treatment of hydrogenlike atoms, $\zeta=Z/a_0$ is given by the core charge number
$Z$. Here, we take $\zeta=1/a_0$. Using
\begin{equation}
Y_l^m(\theta, \phi)=\sqrt{\frac{2l+1}{4\pi}\frac{(l-m)!}{(l+m)!}}P_l^m(\cos \theta)
e^{im\phi} \, ,
\end{equation}
[where $P_l^m(x)$ are the associated Legendre functions], it is easy to verify that the
$\phi$ integration gives $\pi
\delta_{m,m'}$. For the $\theta$ integration, we substitute $x=\cos \theta$. Applying the
relation
\cite{Gradstein}
\begin{equation}
x(2l+1)P_l^m(x)=(l-m+1)P_{l+1}^m(x)+(l+m)P_{l-1}^m(x)
\end{equation}
two times and using the normalization
\begin{equation}
\int_{-1}^1 P_l^m(x)
P_{l'}^m(x)dx=\frac{2}{2l+1}\frac{(l+m)!}{(l-m)!}\delta_{ll'} \, ,
\end{equation}
it can be shown that
\begin{eqnarray}
\int_{-1}^1P_l^m(x) P_{l'}^m (x) x^2 dx \nonumber \\ = \left\{ \begin{array}{lr}
\frac{4l^2-1+(l+m)(4l^2+4m-3)}{(2l-1)(2l+1)^2(2l+3)}\frac{2(l+m)!}{(l-m)!} \, , &
l'=l \\ & \\
\frac{2}{(2l'-1)(2l'+1)(2l'+3)}\frac{(l+m+1 \pm 1)!}{(l-m-1 \pm 1)!} \, , & l'=l\pm 2
\end{array} \right.
\end{eqnarray}
and zero otherwise. That means, the matrix element is nonzero only for $l=l'$ or $l=l'
\pm 2$. However, for $l=l'$, the radial integral is nonzero only for $n=n'$; that term is
not needed for this calculation. Combining these results leads to the more simple
expression for the correction of the Born-Mayer potential
\begin{eqnarray}
\delta \Phi_{BM}(\vec R) & = & 2 \, \mbox{const} \sum_{n'=3}^\infty \frac{<nlm|V|n'l+2 m>}{E_n-E_{n'}} \nonumber \\
 & & \times  |\psi_{nlm}^{(0)}(\vec
R)\psi_{n'\, l+2\, m}^{(0)}(\vec R)|
\end{eqnarray}
(for $n=1,2$, $0\leq l<n$, and $|m| \leq l$). Formally, the series should be summed up to
infinity. However, the terms decay rapidly with increasing $n'$. The correction of the
Born-Mayer potential due to Lorentz violation can thus be obtained and compared to the
correction of the Coulomb potential due to the same Lorentz violation (Fig.
\ref{potentialvergleich}). Here, $\vec R$ was assumed parallel to the $z$ direction. If $\vec R$
is not perpendicular to the $z$ direction, the correction to the Born-Mayer potential is
smaller. From this calculation, it can be concluded that the Lorentz-violating
modification of the repulsive potential $\delta \Phi_{BM}$ is negligible compared to the
modification of the Coulomb potential, at least for the $n=1$ and 2 states that have been
considered in this simple model. This is due to the small magnitude of the matrix
elements $<nlm|V|n'l'm'>$, which are also small for high $n$ and $n'$ (they even decrease
with increasing $n$ and $n'$). So, although for $n \geq 3$, the perturbation theory for
degenerate states should be applied, this will not dramatically change the magnitude of
$\delta \Phi_{BM}$ and thus our conclusion that $\delta \Phi_{BM}$ is negligible remains
valid.


\begin{references}
\bibitem{comment2} Other parameters $(k_{AF})^\kappa$ are strongly bounded by astrophysical polarization
measurements. We assume $(k_{AF})^\kappa=0$ here.
\bibitem{comment1} Lorentz violation might also influence the boundary conditions for the electromagnetic
fields inside the cavity, adding a phase shift $\phi_b$ to radiation reflected at the
boundary. This results in a frequency shift $\delta_b \nu/\nu = \phi_b \lambda /(2\pi L)$
($\lambda$ is the wavelength inside the cavity), which would have to be added to
(\ref{deltac}). However, it is negligible, at least in optical cavities, where
$\lambda/L$ is of order $10^{-3 \ldots 6}$.
\end{references}
\end{document}